\documentclass[twocolumn,showpacs,preprintnumbers,amsmath,amssymb]{revtex4}
\usepackage{graphicx}
\usepackage{dcolumn}
\usepackage{bm}

\begin{document}

\preprint{APS/123-QED}
\title{Multiple Reflections and Diffuse Scattering in Bragg Scattering at Optical Lattices}

\author{S. Slama}
\author{C. von Cube}
\author{M. Kohler}
\author{C. Zimmermann}
\author{Ph.W. Courteille}
\affiliation{Physikalisches Institut, Eberhard-Karls-Universit\"at T\"ubingen,
\\Auf der Morgenstelle 14, D-72076 T\"ubingen, Germany}

\date{\today}

\begin{abstract}
We study Bragg scattering at $1D$ atomic lattices. Cold atoms are confined by optical dipole forces at the antinodes of a standing wave 
generated inside a laser-driven cavity. The atoms arrange themselves into an array of lens-shaped layers located at the antinodes of 
the standing wave. Light incident on this array at a well-defined angle is partially Bragg-reflected. We measure reflectivities as high 
as $30\%$. In contrast to a previous experiment devoted to the thin grating limit [S.~Slama, \textit{et al.}, Phys. Rev. Lett. 
\textbf{94}, 193901 (2005)] we now investigate the thick grating limit characterized by multiple reflections of the light beam between 
the atomic layers. In principle multiple reflections give rise to a photonic stop band, which manifests itself in the Bragg diffraction 
spectra as asymmetries and minima due to destructive interference between different reflection paths. We show that close to resonance 
however disorder favors diffuse scattering, hinders coherent multiple scattering and impedes the characteristic suppression of 
spontaneous emission inside a photonic band gap. 
\end{abstract}

\pacs{42.50.Vk, 42.55.-f, 42.60.Lh, 34.50.-s}

\maketitle

\section{Introduction}
\label{SecIntroduction}

The idea of realizing photonic band gaps (PBG) in optical lattices, i.e.~periodic arrays of cold atomic clouds confined inside standing 
light waves, has been published in 1995 by Deutsch \textit{et al.} \cite{Deutsch95}. A first step towards an experimental observation of 
this phenomenon was made by showing that lattices of atomic gases can give rise to Bragg scattering in the very same way as X-rays are 
scattered in structure analyses of solid crystals \cite{Wollan32} or molecules \cite{Doucand87}. This demonstration has been given 
by G. Birkl \textit{et al.} and  M. Weidem\"uller \textit{et al.} \cite{Birkl95,Weidemuller95} using near-resonant optical lattices. 

%Conservative lattices

In resonant lattices the optical trapping potential provides an efficient cooling mechanism for the atomic cloud. This cooling is 
important, because it balances heating due to inelastic scattering processes, which destroy the periodic order and decrease the Bragg 
scattering efficiency. Cooling is absent in conservative optical lattices tuned far from atomic resonances. On the other hand, 
conservative lattice potentials are interesting in view of their perspectives to mimic solid state physics. For example, Mott 
insulator phase transitions in degenerate atomic quantum gases have been observed \cite{Greiner02}, and fermionic gases confined in 
optical lattices are expected to exhibit novel quantum phases involving high-temperature superfluidity. Those phases may constitute 
useful toy models for superconductivity in high-$T_c$ cuprates \cite{Hofstetter02}. Bragg diffraction could represent a novel and 
powerful tool for sensitively probing the properties of such optical crystals provided the destructive influence of resonant probe light 
absorption is mastered. 

%Photonic cristals

PBGs are today extensively studied in crystals and fibers. Dielectric materials offer the possibility of realizing complex periodic 
structures in three dimensions alternating high and low index of refraction domains. Those structures, called photonic crystals, can 
exhibit ranges of frequencies known as photonic band gaps for which the propagation of electromagnetic waves is classically forbidden 
in certain directions \cite{John83}. Tailoring of the density of states for the electromagnetic modes allows for controlling fundamental 
atom-radiation interactions in solid state environments and even to suppress vacuum fluctuations. The hallmarks of a PBG are the 
inhibition of spontaneous emission, an effect that has been observed with optical cavities \cite{Heinzen87} and the possibility of 
Anderson localization of light by point defects added to the photonic band gap material. 

Although impressive progress has been made \cite{Woldeyohannes03} in fabricating photonic crystals, they suffer from fundamental 
difficulties in guaranteeing the required fidelity over long ranges \cite{Koenderink03} due to fluctuations in position and size of the 
building blocks. This disorder perturbs those properties of photonic crystals based on global interference: It reduces the Bragg 
reflectivity, extinguishes the transmitted light, and ultimately destroys the photonic band gap. On the other hand, optical lattices 
exhibit an intrinsically perfect periodicity. Local disorder introduced by thermal fluctuations in the atomic density distribution 
at each lattice site reduces the value of the Debye-Waller factor \cite{Weidemuller98}, but does not affect the quality of the 
long-range order. 

%Photonic band gaps in optical lattices

To observe photonic band gaps with optical lattices, one must reach the \textit{thick grating regime}. However, all Bragg scattering 
experiments on optical lattices have so far \cite{Birkl95,Weidemuller95,Slama05} been performed in the \textit{thin grating regime}, 
where the lattice's optical density is so low that multiple light scattering events are rare. Bragg scattering at thin lattices is 
understood as resulting from constructive interference of the Rayleigh-scattered radiation pattern emitted by periodically arranged 
point-like sources. In this regime, the reflection coefficient of the lattice turns out to be nearly real, phase shifts are negligible, 
and spectral lineshapes are symmetric. In particular, since the scattering takes place as a local process, which means that the light is 
scattered by individual atoms, the \textit{atomic positions do not shape the absorption spectrum} \cite{Note01}. 

In contrast, the thick grating regime is characterized by multiple reflections of the incident light between the stacked atomic layers 
(Bragg planes). The interference between the light reflected from or transmitted through the layers gives rise to stopping bands for 
certain light frequencies or irradiation angles. In this regime, absorption can generally be neglected, but large phase shifts occur, 
and the lineshapes are asymmetric. Multiple beam interference globalizes the scattering process. 

\bigskip

%This work

In this work we study a one-dimensional optical lattice consisting of a standing light wave filled with trapped atoms. The atoms arrange 
themselves into a linear array of lens-shaped clouds aligned along their symmetry axis. The clouds have a finite radial extent and are 
centered at the locations of the antinodes. We show that for this configuration the thick grating regime is within experimental reach. 
In fact, which regime is realized in experiment depends on the available effective number of scattering layers. The number of populated 
antinodes sets an upper limit. However the finite radial extent of the layers also limits the number of multiple reflections when the 
angle of incidence for the Bragg light is large. Therefore, to get into the thick grating regime, we have modified the setup of a 
previous experiment \cite{Slama05} in order to reduce the Bragg angle and increase the radial extent of the Bragg layers. With this setup 
we encounter a new limitation: When the probe laser is tuned close to an atomic resonance, detrimental absorption due to disordered atoms 
reduces the effective number of layers involved in multiple scattering and dominates the spectra. Off-resonance, in contrast, the weak 
Bragg scattering efficiency brings us into the thin grating regime. Hence, it is important to identify clear signatures of multiple 
scattering in optical lattices and to develop sensitive tools for their detection. 

In some respects, we may understand the $1D$ optical lattice as a dielectric mirror with layers made of a dilute atomic gas. On the other 
hand, as shown in Ref.~\cite{Slama05b}, the $1D$ optical lattice shares the peculiarities of a linear array of point-like scatterers. 

Several authors \cite{Birkl95,Coevorden96,Weidemuller98} suggested optical lattices for the creation of $3D$ photonic band gaps. However, 
a consequence of the narrow width of the atomic resonance is that \textit{Bragg scattering at gaseous lattices is intrinsically 
one-dimensional}. This holds also for $2D$ and $3D$ geometries of optical lattices, so that the results of our investigations apply to 
all kinds of lattice configurations. The low dimensionality of the scattering problem thus compromises the realization of true $3D$ 
photonic band gaps with optical lattices. 

\bigskip

%Organization

We organized this paper as follows: In section~\ref{SecExp} we present our experimental setup, show examples of typical spectra, and 
discuss how diffuse scattering interferes with multiple reflections. To qualitatively understand the spectra a transfer matrix model is 
developed in section~\ref{SecTheory}. Based on Ref.~\cite{Deutsch95} the model is extented to comply with partially disordered lattices 
and inhomogeneous Stark shifts. The model provides a simple picture for the observed spectra allowing for a discrimination between 
diffuse and multiple scattering. It also describes the expected suppression of spontaneous emission and yields a quantitative prediction 
of the lattice's reflection and transmission as a function of experimental parameters. Section~\ref{SecObs} presents our observations 
and discusses them in terms of multiple reflections and diffuse scattering. Since both effects give rise to similar signatures in the 
reflection spectra, they are difficult to separate, in particular in the presence of experimental imperfections. As summarized in 
section~\ref{SecConclude}, despite the fact that some signatures strongly suggest the concurrence of multiple reflections, it seems 
actually beyond reach to see suppression of spontaneous emission due to a reduced density of optical states.

\section{Experiment}
\label{SecExp}

\subsection{Setup}
\label{SecExpSetup}

\begin{figure}[ht]
		\centerline{\scalebox{0.41}{\includegraphics{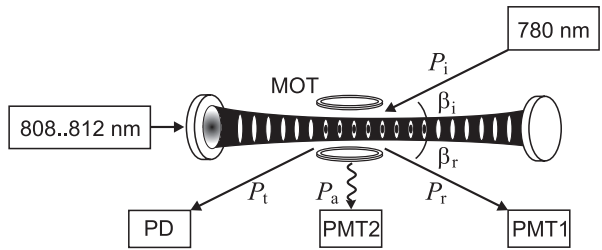}}}\caption{
			The experimental setup consists of a cavity pumped with a titanium-sapphire laser at $808$ - $812~$nm and a diode laser at 
			$780~$nm for Bragg scattering. About 10000 antinodes of the cavity mode are loaded with $^{85}$Rb atoms from a MOT. Observed 
			are the reflected (PMT1), the transmitted (PD) and the absorbed (PMT2) light powers.}
		\label{Fig1}
		\end{figure}

The optical layout of our experiment shown in Fig.~\ref{Fig1} is identical to the one presented in Ref.~\cite{Slama05b}. It consists of 
an optical cavity and a setup for Bragg scattering. The light of a titanium-sapphire laser operating at 
$\lambda_{\mathrm{dip}}=808$ - $812~$nm, which is red-detuned with respect to the rubidium $D_1$ line, is coupled and phase-locked to 
the cavity. The standing wave which builds up inside the cavity has a periodicity of 
$\frac{1}{2}\lambda_{\mathrm{dip}}=\pi/k_{\mathrm{dip}}$. The beam diameter at the center of the cavity is $w_{\mathrm{dip}}=220~\mu$m. 
The intracavity light power is $P_{\mathrm{cav}}=5~$W. Between $N=10^5$ and $10^7$ $^{85}$Rb atoms are loaded from a standard 
magneto-optical trap (MOT) into the standing wave. About $10000$ antinodes are filled with atoms. Typically the temperature of the cloud 
is on the order of few $100~\mu$K. In earlier experiments \cite{Slama05b} we found that the temperature of the cloud tends to adopt a 
fixed ratio with the depth of the dipole trap, $k_{\mathrm{B}}T\approx0.4U_0$ \cite{Note02}. Therefore, the spatial distribution of the 
atoms does not vary much with the potential depth. From this we derive the \textit{rms}-size of an individual atomic cloud along the 
cavity axis, $2\sigma_z=\frac{1}{\pi}\lambda_{\mathrm{dip}}\sqrt{k_\mathrm{B}T/2U_0}\approx115~$nm in the harmonic approximation of the 
trapping potential. The radial size is $2\sigma_r=w_{\mathrm{dip}}\sqrt{k_{\mathrm{B}}T/U_0}\approx140~\mu$m and the mean density lies 
between $n=3\times10^9$ and $3\times10^{11}~$cm$^{-3}$. For the present setup we estimate a Debye-Waller factor of 
$f_{\mathrm{DW}}=e^{-2k_{\mathrm{dip}}^2\sigma_z^2}\approx 0.2$ \cite{Slama05b}. 

The light used to probe the Bragg resonance is generated with a near infrared laser diode operating at $\lambda_{\mathrm{brg}}=780~$nm. 
The laser light is passed through an acousto-optic modulator (AOM) and an optical fiber and collimated to a beam waist of 
$w_{\mathrm{brg}}=800~\mu$m before crossing the dipole trap standing wave under an angle of 
$\beta_{\mathrm{i}}\simeq\arccos(\lambda_{\mathrm{brg}}/\lambda_{\mathrm{dip}})$, which at $\lambda_{\mathrm{dip}}=810~$nm is about 
$15.6^\circ$. The incident laser power, $P_{\mathrm{i}}=30~\mu$W, is well below saturation. Some time after loading the atoms into the 
standing wave the probe beam is switched on and frequency-ramped across the rubidium $D_2$ resonance. The light power reflected from the 
atoms $P_{\mathrm{r}}$ is detected under the angle $\beta_{\mathrm{s}}=-\beta_{\mathrm{i}}$ with a photomultiplier (PMT1). The 
transmitted light power $P_{\mathrm{t}}$ is recorded with a photodiode (PD), and the isotropically scattered power $P_{\mathrm{a}}$ is 
detected by collecting the light emission into a solid angle of $0.05~$sr orthogonal to the incident probe beam with a second 
photomultiplier (PMT2). 

To obtain Bragg reflection the angle of incidence of the probe laser has to be matched to the lattice constant. Experimental fine tuning 
of the Bragg condition is however easier by varying the wavelength of the lattice laser, $\Delta\lambda_{\mathrm{dip}}\equiv\lambda_
{\mathrm{dip}}-\lambda_{\mathrm{brg}}/\cos\beta_{\mathrm{i}}$, while the angle of incidence is kept fixed.

\subsection{Bragg spectra}
\label{SecExpBragg}

The experimentally accessible quantities are the reflected, transmitted and absorbed light powers. We take simultaneous spectra of these 
quantities by ramping the probe laser frequency across the Bragg resonance. In order to compare with calculations we are interested in 
the reflection, transmission and absorption coefficients $R,T$, and $A$. A direct comparison is complicated by the fact that the probe 
beam cross section is larger than the size of the atomic cloud, so that only a fraction $\eta\approx16\%$ of the incident power 
$P_{\mathrm{i}}$ really overlaps with the atomic cloud \cite{Note03}, 
	\begin{align}\label{EqExp01}
	P_{\mathrm{r}} & =R\eta P_{\mathrm{i}}~,\\
	P_{\mathrm{t}} & =T\eta P_{\mathrm{i}}+(1-\eta)P_{\mathrm{i}}\nonumber\\
	P_{\mathrm{a}} & =A\eta P_{\mathrm{i}}\nonumber~.
	\end{align}
Therefore the energy conservation requirement, $P_{\mathrm{i}}=P_{\mathrm{r}}+P_{\mathrm{t}}+P_{\mathrm{a}}$, implies $R+T+A=1$. 

\begin{figure}[ht]
		\centerline{\scalebox{0.48}{\includegraphics{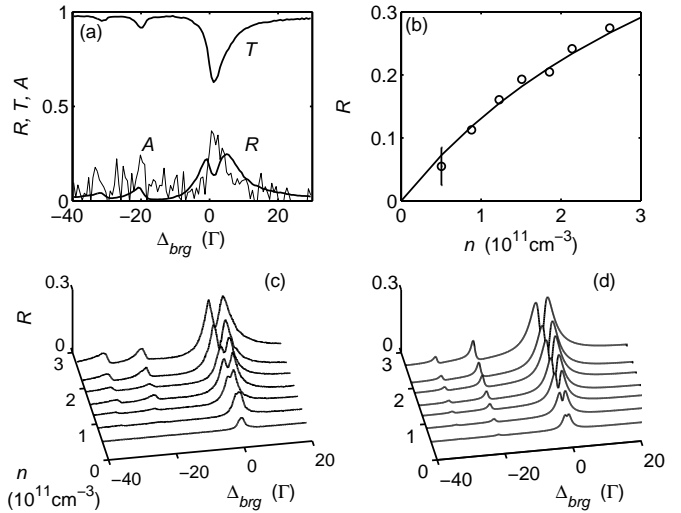}}}\caption{
			\textbf{(a)} Reflection (lower thick trace), transmission (upper trace) and absorption spectra (noisy thin trace) of Bragg 
			scattering. 
			\textbf{(b)} Dependence of the absolute maximum of the reflection spectra on the atom number. The maximum Bragg diffraction 
			efficiency saturates for large atom numbers. The theoretical curve assumes $N_{\mathrm{s}}=400$ scattering layers and a 
			Debye-Waller factor of $f_{\mathrm{DW}}=0.2$. 
			\textbf{(c)} Measured Bragg spectra for various atomic densities. The small peaks at $\Delta_{\mathrm{brg}}=-31\Gamma$ and 
			$-20\Gamma$ corresponds to transitions from the ground state hyperfine level $F=3$ to the excited state levels $F'=2,3$, the 
			broad peak around $\Delta_{\mathrm{brg}}=0$ to the level $F'=4$. 
			For the calculations \textbf{(d)} the optical cross sections at the various hyperfine transitions have been weighted according 
			to their relative oscillator strength.}
		\label{Fig2}
		\end{figure}

Fig.~\ref{Fig2}(a) shows reflection, transmission and absorption spectra of the Bragg resonance obtained by ramping the detuning of the 
probe laser $\Delta_{brg}$ from the $D_2$ resonance. The resonance linewidth is $\Gamma/2\pi=6~$MHz. Note the high Bragg reflection 
efficiency of more than $30\%$, which is more than two orders of magnitude higher than in any previous measurement on optical lattices. 

The scattering efficiency for standard Bragg diffraction depends quadratically on the atom number. This dependency has been observed in 
Ref.~\cite{Slama05} in the thin grating regime. The measurements performed with the present apparatus exhibit a different behavior. As 
seen in Fig.~\ref{Fig2}(b) at large atom densities the scattering efficiency seems to saturate. Also the shape of the reflection signal 
depends critically on the atom number. At very low atom number we find a Lorentzian lineshape, whose width corresponds to the natural 
linewidth of the $D_2$ line. When the atom number is increased, the linewidth broadens and the peak saturates. At large atom numbers a 
pronounced dip appears at the center of the strongest resonance of the reflection spectrum \cite{Birkl95}, whose contrast increases with 
increasing atomic density [cf. Fig.~\ref{Fig2}(c)]. Theoretical predictions based on transfer matrices calculations presented below 
confirm this behavior [cf. Fig.~\ref{Fig2}(d)]. The main goal of this paper is to explain these observations.

\subsection{Specular versus diffuse scattering}
\label{SecExpSpecular}

The atoms are strongly localized in axial direction at the center of the antinodes. Consequently, the spectrum of the light reflected 
into the Bragg angle is narrowed by the Lamb-Dicke effect, and the axial thermal distribution of the atoms does not broaden the angular 
distribution of the reflected radiation, but increases the background of isotropically distributed diffuse scattering. This behavior is 
known from Bragg scattering of X-rays at solids. In contrast the weak radial confinement of the atoms does broaden the angular 
distribution \cite{Slama05b}. 

The finite radial atomic distribution has however a more important impact on the reflected light. The radial size of the atomic cloud 
determines the length of the probe beam trajectory across the lattice. Thus for thick lattices, when multiple scattering plays a role, 
the maximum number of reflections, i.e.~the effective number of layers, is limited to 
$N_{\mathrm{s}}=2\sigma_r/(\lambda_{\mathrm{dip}}\tan\beta_i)$. In an earlier experiment \cite{Slama05} we have studied Bragg 
scattering at an optical lattice near the $5S_{1/2}$ - $6P_{3/2}$ transition at $420~$nm. The corresponding Bragg angle was $58^{\circ}$ 
yielding $N_{\mathrm{s}}\approx100$. In the present setup we use a linear cavity with a larger mode waist and operate the probe light
near the $D_2$ line. This decreases the Bragg angle to $15.6^\circ$ and increases the effective number of layers to 
$N_{\mathrm{s}}\approx600$. 

\bigskip

At first glance, the saturation behavior observed in Fig.~\ref{Fig2}(b) could be interpreted as an indication for multiple scattering. 
The dips appearing in Fig.~\ref{Fig2}(c) could result from different multiple scattering trajectories of the probe laser along the 
lattice, which destructively interfere for certain values of the incident angle and of the probe light frequency. However, as already 
demonstrated in Ref.~\cite{Birkl95} the situation is more complex. In fact in a thermal lattice, for a small Debye-Waller factor, 
$f_{\mathrm{DW}}\ll 1$, disordered atoms have a dramatic influence on the scattering: 1.~They reduce the number of ordered atoms 
contributing to Bragg scattering. 2.~They absorb and attenuate the incident light, and hence \textit{decrease the penetration depth of 
the probe beam and thus the effective number of layers available to Bragg and in particular to multiple scattering.} 
%3.~The disordered atoms phase shift the probe beam along its trajectory through the atomic cloud. 

A first approach in describing the physical situation may consist in dividing the atomic cloud into two parts: a perfectly ordered 
optical lattice with density $n f_{\mathrm{DW}}$ and a homogenous cloud having the density $n (1-f_{\mathrm{DW}})$. The homogeneous 
cloud limits the penetration depth to $z_{\mathrm{pd}}=[\sigma n (1-f_{\mathrm{DW}})]^{-1}$, which corresponds to a reduced effective 
number of layers $N_{\mathrm{s,pd}}=2z_{\mathrm{pd}}/\lambda$. The effective number of layers thus critically depends on the density of 
disordered atoms and, via the optical cross section $\sigma$, on the detuning $\Delta_{\mathrm{brg}}$. On resonance assuming a density 
$n\approx3\times10^{11}~$cm$^{-3}$ we estimate $N_{\mathrm{s,pd}}\approx37$ \cite{Note04}. As a consequence, we expect a dramatic 
break-down of the Bragg reflection signal close to resonance, where the absorption by the unordered atoms is largest. The frequency 
range for absorption, being on the order of the natural linewidth $\Gamma$, is much smaller than the width of the reflection signal, 
which explains the appearance of a narrow dip in the reflection spectra Fig.~\ref{Fig2}(c). In particular, \textit{signatures of multiple 
scattering will not show up in parameter regimes, where the number of scattering layers is considerably reduced below $N_{\mathrm{s}}$}, 
$N_{\mathrm{s,pd}}\ll N_{\mathrm{s}}$. 

%Also problematic are phase shifts caused by the disordered cloud. Besides introducing asymmetric line shapes, these phase shifts can 
%strongly perturb the signatures of multiple reflections, because multiple reflections are essentially based on the dispersive part of 
%the polarizability. 

Note that the separation of the Bragg reflection into a perfectly ordered contribution and diffuse scattering corresponds to an ansatz 
frequently made in treating disorder and impurities in crystals \cite{Weidemuller98,Coley01}.

\section{Theoretical modeling}
\label{SecTheory}

\subsection{Transfer matrix formalism}
\label{SecTheoryTransfer}

In the following we attempt to provide a more quantitative understanding of the observations by developing a simple theoretical model. 
We start by relating the atomic polarizability $\alpha$ (given in SI-units) to the single-layer reflection coefficient $\zeta$ via 
	\begin{align}\label{EqTheory01}
	\zeta & =-n\delta z~\frac{k_{\mathrm{brg}}}{2}~\frac{\alpha}{\epsilon_0}\\
	& =-n\delta z~\frac{3}{2}\frac{\lambda_{\mathrm{brg}}^2}{2\pi}\frac{1}{i+2\left(\Delta_{brg}-\Delta_F\right)/\Gamma}~,\nonumber
	\end{align}
where $n\delta z$ is the surface density estimated for a homogeneous atomic density $n$ and a layer thickness $\delta z$. To account for 
the presence of several hyperfine transitions at frequency detunings $\Delta_F$ [cf. Fig.~\ref{Fig2}(d)], we build the weighted sum of 
the individual oscillator strengths. 

For the description of the collective influence of the atoms on the incident light, we use a generalization of the transfer matrix 
formalism presented in Ref.~\cite{Deutsch95}. The generalization concerns two major points: First of all, in that reference the probe 
beam was assumed collinear with the optical lattice beams, while in our case the angle of incidence is significantly different from 0. 
In fact the deviation of the chosen angle from the Bragg angle constitutes an additional degree of freedom allowing us to tune frequency 
and quasi-momentum independently. Again, in practice we detune the lattice constant rather than the angle of incidence. A second 
generalization consists in the inclusion of diffuse scattering into the formalism, as detailed in the next section. All theoretical 
lineshapes shown in the figures are obtained from this transfer matrix model. 

We will now supply the basic ingredients of the model skipping the details already reported in Ref.~\cite{Deutsch95}. The in- and 
outgoing field amplitudes of the probe beam at any axial location $z$ of the lattice (the model is one-dimensional) are labeled $E^+(z)$ 
and $E^-(z)$, respectively. Their variation from one location to another is described by a transfer matrix $M$, such that (in the complex 
representation) 
	\begin{align}\label{EqTheory02}
	\left(\begin{array}[c]{cc}
		E^+(z)\\
		E^-(z)
	\end{array}\right)
	=M\left(\begin{array}[c]{cc}
		E^+(0)\\
		E^-(0)
	\end{array}\right)~.
	\end{align}

The procedure consists now in dividing the atomic sample into layers. The transfer matrix for interaction of the probe light with a 
single infinitely narrow layer of the optical lattice characterized by the surface density $n\delta z$ is 
	\begin{align}\label{EqTheory03}
	A_{\zeta} & =\left(\begin{array}[c]{cc}
		1+i\zeta &  i\zeta\\
		-i\zeta & 1-i\zeta
	\end{array}\right)~.
	\end{align}
The transformation of the field amplitudes \textit{between} two such layers separated by $\Delta z$ is described by 
	\begin{align}\label{EqTheory04}
	B_{\Delta z} & =\left(\begin{array}[c]{cc}
		e^{ik_{\mathrm{brg}}\Delta z\cos\beta_{\mathrm{i}}} & 0\\
		0 & e^{-ik_{\mathrm{brg}}\Delta z\cos\beta_{\mathrm{i}}}
	\end{array}\right)~.
	\end{align}

\begin{figure}[ht]
		\centerline{\scalebox{0.48}{\includegraphics{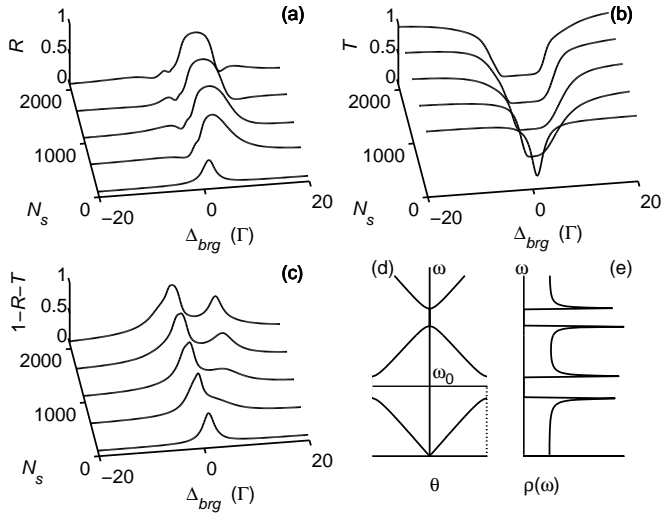}}}\caption{
			The figures \textbf{(a)}, \textbf{(b)}, and \textbf{(c)} show calculated Bragg reflection, transmission and absorption spectra 
			for lattices with various numbers of layers $N_s$. Here we assumed $f_{\mathrm{DW}}=1$ and $n=3\times10^{11}~$cm$^{-3}$. The 
			lattice is slightly detuned from the Bragg condition, $\Delta\lambda_{\mathrm{dip}}=0.8~$nm, so that the lineshapes become 
			asymmetric. \textbf{(d)} Sketch of the photonic dispersion relation. The opening of a band gap for values of the 
			quasi-momentum $\theta$ near the edges of the Brillouin zone indicates a suppression of the density of states $\rho(\omega)$ 
			as drawn in figure~\textbf{(e)}.}
		\label{Fig3}
		\end{figure}

Hence the total transfer matrix for a lattice with $N_{\mathrm{s}}$ layers reads $M=\left(A_{\zeta}B_{\Delta z}\right)^{N_{\mathrm{s}}}$. 
Finally the reflection coefficient $R=|r|^2$ and the transmission coefficient $T=|t|^2$ are calculated via 
	\begin{equation}\label{EqTheory05}
	r=\frac{M_{12}}{M_{22}}~~~~~,~~~~~	t=\frac{1}{M_{22}}~,\\
	\end{equation}
while the phase shift in reflection follows from $\phi=\arctan(\operatorname{Im}r/\operatorname{Re}r)$. If we identify a layer with an 
antinode of the standing wave, $\Delta z=\delta z=\lambda_{\mathrm{dip}}/2$, we obtain from the Eqs.~(\ref{EqTheory05}) Bragg spectra for 
the case of a perfect lattice, such as those shown in Figs.~\ref{Fig3}(a-c). 

\subsection{Sequential density model}
\label{SecTheorySequential}

Instead of separating the atomic cloud into a perfectly ordered lattice and a homogeneous density distribution as proposed in 
Sec.~\ref{SecExpSpecular}, we may subdivide every layer into a number $N_{\mathrm{ss}}$ of sublayers for which we evaluate the transfer 
matrices based on the local density, 
	\begin{align}\label{EqTheory06}
	n_{\mathrm{loc}}(z) & =n\frac{\lambda_{\mathrm{dip}}e^{-U(z)/k_{\mathrm{B}}T}}{2N_{\mathrm{ss}}\int e^{-U(z')/k_{\mathrm{B}}T}dz'}~,
%\\
%	& =n\frac{e^{-(1-2\nu/N_{\mathrm{ss}})^2(\lambda_{\mathrm{dip}}/4)^2/2\sigma_z^2}}{\sum_{\nu'=1}^{N_{\mathrm{ss}}}e^{-(1-2\nu'/
%		N_{\mathrm{ss}})^2(\lambda_{\mathrm{dip}}/4)^2/2\sigma_z^2}}~,\nonumber
	\end{align}
where $U(z)=-U_0\cos k_{\mathrm{dip}}z$ is the trapping potential, or in the harmonic approximation 
$U(z)=U_0 k_{\mathrm{dip}}^2z^2=k_{\mathrm{B}}T z^2/2\sigma_z^2$. Now we recalculate the local single-layer reflection coefficient 
$\zeta_{\mathrm{loc}}$ as in Eq.~(\ref{EqTheory01}) and set up the total transfer matrix, but using the \textit{local} density 
$n_{\mathrm{loc}}(z)$ instead of a homogeneous density $n$ and setting the layer thickness to 
$\delta z\equiv\lambda_{\mathrm{dip}}/2N_{\mathrm{ss}}$, 
	\begin{equation}\label{EqTheory07}
	M=\left(\prod_{\nu=1}^{N_{\mathrm{ss}}} A_{\zeta_{\mathrm{loc}}}(\tfrac{2\nu\delta z}{\lambda_{\mathrm{dip}}})~B_{\delta z}\right)
		^{N_{\mathrm{s}}}~.
	\end{equation}
 		\begin{figure}[ht]
		\centerline{\scalebox{0.48}{\includegraphics{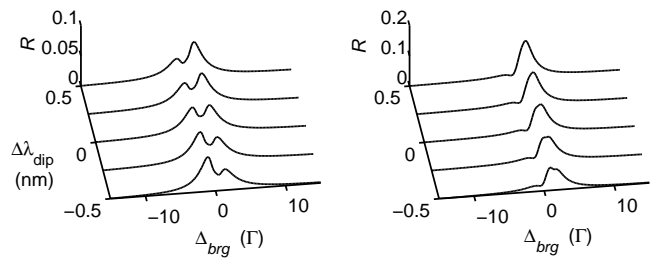}}}\caption{
			Calculated Bragg reflection spectra for various detunings of the lattice constant from the Bragg condition \textbf{(a)} without 
			and \textbf{(b)} with Stark-shift. We assumed $n=3\times10^{11}~$cm$^{-3}$, $N_{\mathrm{s}}=200$ and $N_{\mathrm{ss}}=20$. 
			At a potential depth $U_0=500~\mu$K we assume $k_{\mathrm{B}}T\approx0.4U_0$.}
		\label{Fig4}
		\end{figure}

\bigskip

At finite temperatures the atoms are distributed over the optical potential and thus experience individual dynamical Stark shifts of 
their resonances varying with the atoms' locations. This leads to serious inhomogeneous broadening of the Bragg spectra as shown in 
Ref.~\cite{Slama05}. Cold atoms which concentrate at the antinodes of the standing wave potential are Stark-shifted by a large amount and 
form the blue edge of the line profile. The hot part of the cloud sees on average a shallower potential and forms the red tail of the 
profile. A possible approach to describe the line broadening consists in building the convolution of the spectra calculated from 
Eqs.~(\ref{EqTheory05}) and (\ref{EqTheory07}) with the probability density of finding an atom at a given potential energy. However this 
approach does not account for the fact that in the thick grating limit cold (predominantly ordered) and hot (mostly disordered) atoms 
yield qualitatively different lineshapes of the Bragg reflection signal. In other words, the convolution procedure is incompatible with 
the fact that the contributions of cold and hot atoms to the Bragg-scattered light depends on the penetration depth, which itself varies 
with the frequency detuning of the probe beam. 

Fortunately, the local Stark shift is easily included in the sequential densities model via the substitution 
	\begin{equation}\label{EqTheory08}
	\Delta_{\mathrm{brg}}\rightarrow\Delta_{\mathrm{brg}}-U(z)~.
	\end{equation}
The figures~\ref{Fig4}(a,b) show calculated reflection spectra without and with Stark shifts for various detunings of the lattice 
constant from the Bragg condition. As in Figs.~\ref{Fig2}(b,c) the dip in the spectra corresponds to a joint impact of diffuse and 
multiple scattering and will be discussed in more detail in Sec.~\ref{SecObsSignatures}. Here we just point out that Stark broadening 
obviously induces pronounced asymmetries with respect to $\Delta\lambda_{\mathrm{dip}}$. 

We verified that in the limit of a perfect lattice, obtained for $T\rightarrow0$, the line broadening disappears. The lines are just 
blue-shifted by an amount $U_0/\hbar$. In the thin grating regime, obtained for small densities $n<10^{11}~$cm$^{-3}$, the convolution 
approach and the sequential densities approach yield identical spectra. 
		\begin{figure}[ht]
		\centerline{\scalebox{0.48}{\includegraphics{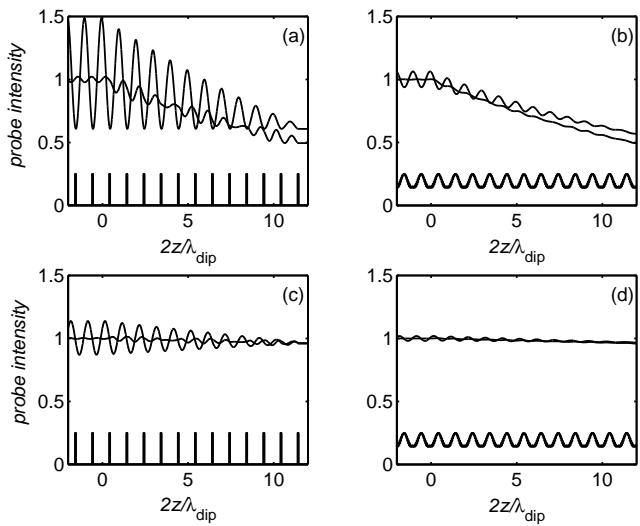}}}\caption{
			\textbf{(a)} Evolution of the light intensity across the optical lattice normalized to the incident probe beam intensity for 
			$\Delta\lambda_{\mathrm{dip}}=0$ (large amplitude oscillations) and $\Delta\lambda_{\mathrm{dip}}=200~$nm (small amplitude 
			oscillations). The last detuning has been chosen to be extreme in order to visualize the rapid decay of the intensity even for 
			a low number of layers $N_{\mathrm{s}}=10$. The probe was tuned to resonance, $\Delta_{\mathrm{brg}}=0$. The spikes at the 
			bottom of 	the pictures indicate the maxima of the atomic density distribution assumed to describe a perfect lattice, 
			$f_{\mathrm{DW}}=1$. 
			\textbf{(b)} Same as (a) but with $f_{\mathrm{DW}}\simeq0.03$. The lowest curve shows the axial atomic density distribution. 
			\textbf{(c,d)} Same as (a) and (b) respectively, but now the probe laser has been detuned to $\Delta_{\mathrm{brg}}=\Gamma$.}
		\label{Fig5}
		\end{figure}

\subsection{Suppression of absorption}
\label{SecTheorySupress}

The hallmark of $3D$ photonic crystals is the suppression of spontaneous emission. The models describing the propagation of light 
inside photonic crystals assign this suppression to a reduction of the density of optical modes available for spontaneous decay 
\cite{John83}. In fact the frequency dependence of the density of states, sketched in Fig.~\ref{Fig3}(e), is characterized by forbidden 
bands. As pointed out in Ref.~\cite{Deutsch95}, $1D$ optical lattices exhibit similar band gaps. This can be seen in 
Fig.~\ref{Fig3}(a-c). For large numbers of scattering layers, $N_{\mathrm{s}}\gtrsim 1000$, the lattice gets opaque. The transmission 
vanishes over a large range of frequency detunings $\Delta_{\mathrm{brg}}$, while the reflection is close to unity. Even more interesting 
is the feature that the absorption spectrum $1-R-T$ splits into two peaks, and the absorption vanishes in the center of the band gap. 

At first glance this may seem surprising. Since the geometry of our lattice is $1D$, we would not expect a noticeable modification of 
the density of decay modes. This is similar to the inhibition of spontaneous emission inside linear optical cavities \cite{Heinzen87}: 
An excited atom may decay into all the transverse modes leading out of the cavity. If the cavity only covers a small solid angle of the 
radiating atom, its spontaneous decay rate will only be reduced by a small amount. 

The reduction of the absorption inside a band gap of an optical lattice has a different origin. A deeper understanding of the system is 
gained by calculating the progression of the probe light intensity along the optical lattice using the above transfer matrix formalism 
under various conditions (cf. Fig.~\ref{Fig5}). First of all, we find that the standing wave formed by the incident probe beam and the 
Bragg-reflected light adjusts its phase such that its \textit{nodes coincide with the atomic layers}. In that way absorption is 
minimized. If the length of the lattice is finite, the contrast of the standing wave is smaller than 1, i.e.~the probe light intensity 
at the locations of the atomic layers does not vanish. Hence a finite absorption subsists even, when the lattice is perfect and the Bragg 
condition fulfilled. 
		\begin{figure}[ht]
		\centerline{\scalebox{0.48}{\includegraphics{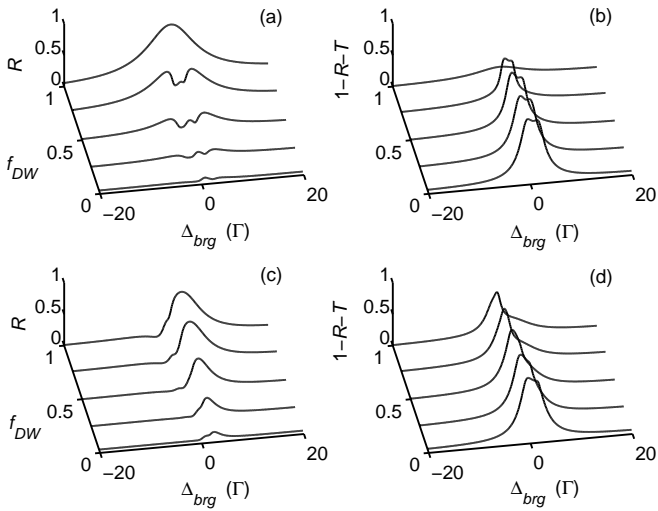}}}\caption{
			Reflection \textbf{(a)} and absorption spectra \textbf{(b)}, calculated for various values of the Debye-Waller factor 
			$f_{\mathrm{DW}}$. The other parameters are $\Delta\lambda_{\mathrm{dip}}=0$, $n=3\times 10^{11}~$cm$^{-3}$, and 
			$N_{\mathrm{s}}=600$. \textbf{(c,d)} Same as (a) and (b), but with $\Delta\lambda_{\mathrm{dip}}=0.8~$nm.} 
		\label{Fig6}
		\end{figure}

Let us now study the response of the probe standing wave to variations of experimental parameters. Fig.~\ref{Fig5}(a) compares the cases 
of an aligned and a misaligned angle of incidence with respect to the Bragg angle. For a misaligned angle the periodicity of the probe 
standing wave does not coincide with the lattice constant, which results in a displacement of the nodes from the atomic layers and hence 
in enhanced absorption. The contrast of the probe wave is smaller than for an aligned angle. However, as the probe light penetrates 
deeper into the lattice, the displacement gets smaller and the probe wave contrast adopts the value of the aligned case. 

The curves shown in Fig.~\ref{Fig5}(a) assume a perfectly ordered lattice. In the presence of disordered atoms, i.e.~atoms which are 
not confined to the locations of the probe beam intensity minima, absorption can take place anywhere. We thus obtain an additional 
background of absorption, which in the extreme case of strong disorder leads to a fast exponential decay according to the Lambert-Beer 
law. The exponential curves in Fig.~\ref{Fig5}(b) are obtained under the same conditions as in (a), but with a finite Debye-Waller 
factor, $f_{\mathrm{DW}}=0.03$. Finally when the probe beam is tuned off resonance, the absorption is smaller and the penetration 
depth is drastically increased. This is shown in Figs.~\ref{Fig5}(c-d), which corresponds to (a-b) respectively, but with a probe laser 
detuning set to $\Delta_{\mathrm{brg}}=\Gamma$. 

\bigskip

In summary, absorption occurs for two reasons: Either the Bragg angle is mismatched, or the atoms are disordered. This is illustrated by 
calculations of reflection and absorption spectra shown in Fig.~\ref{Fig6}. For example, the spectrum in Fig.~\ref{Fig6}(d) corresponding 
to $f_{\mathrm{DW}}=1$ shows an absorption peak despite perfect atomic ordering, because the Bragg angle is not matched; and the 
spectra in Fig.~\ref{Fig6}(b) corresponding to $f_{\mathrm{DW}}<1$ show finite absorption although the Bragg angle is matched. Reduced 
absorption is only observed for a perfect lattice \textit{and} a matched angle of incidence for the probe beam [see curve corresponding 
to $f_{\mathrm{DW}}=1$ in Fig.~\ref{Fig6}(b)].

\section{Observations}
\label{SecObs}

\subsection{Signatures of multiple scattering}
\label{SecObsSignatures}

The appearance of signatures for multiple scattering depends very much on the available effective numbers of layers. As we have seen in 
Sec.~\ref{SecExpSpecular} diffuse scattering drastically reduces the number of layers from $N_{\mathrm{s}}=600$ to 
$N_{\mathrm{s,pd}}=37$, which is clearly insufficient to produce dips in the reflection spectrum corresponding to destructive 
interference of different reflection paths. Fig.\ref{Fig3}(a) reveals that for our atomic densities several hundred layers are necessary 
to significantly distort the spectrum near the photonic band edge. 
		\begin{figure}[ht]
		\centerline{\scalebox{0.48}{\includegraphics{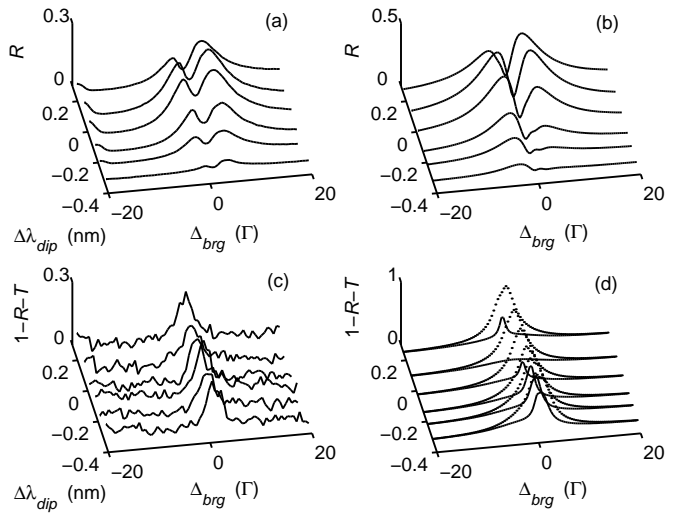}}}\caption{
			Measured \textbf{(a)} and calculated \textbf{(b)} Bragg reflection spectra for various dipole laser frequencies. 
			Measured \textbf{(c)} and calculated \textbf{(d)} absorption. The calculations are based on $n=3\times10^{11}~$cm$^{-3}$, 
				$N_{\mathrm{s}}=200$, and $f_{\mathrm{DW}}=0.7$ (dotted lines) and $f_{\mathrm{DW}}=1$ (solid lines).} 
		\label{Fig7}
		\end{figure}

In order to observe signatures of a band edge, the angle of incidence of the probe beam must be chosen in such a way that they appear 
outside the detuning range, where diffuse scattering is dominant. In fact already at $|\Delta_{\mathrm{brg}}|>3\Gamma$ the penetration 
depth allows for an effective number of layers larger than $600$. 

Fig.~\ref{Fig7}(a) shows Bragg reflection spectra measured for various lattice constants. Fig.~\ref{Fig7}(c) shows simultaneously 
recorded absorption spectra. Figs.~\ref{Fig7}(b,d) represent corresponding calculations. Although the reflection spectra are in 
qualitative agreement, there is a striking difference. While for the measured spectra the dip is always close to the line center, the 
symmetry of the calculated spectra changes when going from negative to positive $\Delta\lambda_{\mathrm{dip}}$. An explanation for 
this is given in Sec.~\ref{SecObsMiscallaneous}. 

The amount of absorption measured in the experiment does not depend much on $\lambda_{\mathrm{dip}}$. This is confirmed by the 
simulations [dotted lines in Fig.~\ref{Fig7}(d)]. The reason for this is diffuse scattering due to atomic disorder as discussed in 
Sec.~\ref{SecTheorySupress} \cite{Note05,Note06}. For comparison we have also plotted in Fig.~\ref{Fig7}(d) the spectra calculated for 
zero temperature (solid lines). The effect of absorption reduction expected for $f_{\mathrm{DW}}=1$ is not observed in experiment. 

\begin{figure}[ht]
		\centerline{\scalebox{0.48}{\includegraphics{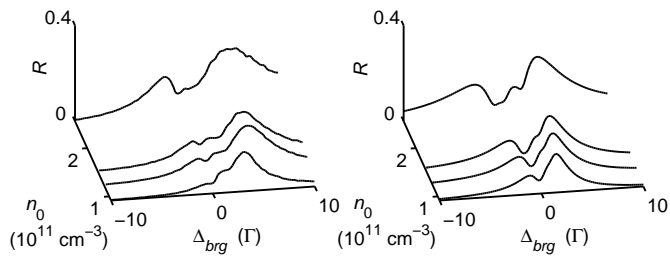}}}\caption{
			Occurrence of double dips in the reflection spectra. Measurements are shown in \textbf{(a)}. The calculations \textbf{(b)} 
			assume $\Delta\lambda_{\mathrm{dip}}=0.08~$nm, $n=3\times10^{11}~$cm$^{-3}$, $N_{\mathrm{s}}=520$, $U_0=0.6\Gamma$, and 
			$T=90~\mu$K.} 
		\label{Fig8}
		\end{figure}

Fig.~\ref{Fig8}(a) shows a set of spectra recorded for various atomic densities. Fig.~\ref{Fig8}(b) shows the corresponding calculations. 
The spectra exhibit a well-developed splitting of the dip structure. This splitting can not be explained by diffuse scattering alone, 
which means that multiple reflections must play a role. 

The number of free parameters is large, therefore it is difficult to pin down the precise value of the experimental parameters by 
fitting. However, there is no realistic parameter regime for which our theoretical model predicts splittings of the dip without the 
assumption of multiple scattering. Hence, we consider the double dip feature of Fig.~\ref{Fig8} as the first indication for the 
existence of a $1D$ photonic band gap in an optical lattice.

\subsection{Experimental side effects}
\label{SecObsMiscallaneous}

Various experimental deficiencies can make the quantitative interpretation of the spectra difficult. First of all, when the probe laser 
is detuned from resonance, its light will be bent by refraction when it enters the optically thick cloud of atoms. On one hand, this 
slightly modifies the angle of incidence and thus its deviation from the Bragg angle. Since this deviation depends on the probe beam 
detuning, the parameters $\Delta\lambda_{\mathrm{dip}}$ and $\Delta_{\mathrm{brg}}$ are intertwined in a complicated way. We have 
measured the reflection angle as a function of $\Delta_{\mathrm{brg}}$ and found variations up to $0.1^{\circ}$, which corresponds to 
$\Delta\lambda_{\mathrm{dip}}=0.4~$nm. Additionally, the optically thick cloud focuses or defocused the incident beam depending on its 
detuning \cite{Note07}. 

A second important issue is the impact of the finite radial extent of the atomic layers on the reflection angle. In Ref.~\cite{Slama05b} 
we have shown that, although the lattice consists of a stack of two-dimensional traps with an aspect ratio smaller than 
$\sigma_z/\sigma_r\approx 10^{-3}$, with respect to Bragg scattering it behaves more like a chain of point-like scatterers, than like a 
dielectric mirror. Thus the reflection angle is not equal to the angle of incidence, but adjusts itself in order to fulfill the Bragg 
condition. This self-adjustment of the Bragg condition impedes a controlled tuning of $\Delta\lambda_{\mathrm{dip}}$ and explains, why 
the symmetry of the traces in Fig.~\ref{Fig7}(a) does not change when the lattice constant is varied. 

Furthermore, the transfer matrix model is purely one-dimensional. It assumes not only radially infinite atomic layers, but also a 
homogeneous density distribution. In reality the radial density distribution is rather Gaussian, which implies a variation of the 
penetration depth with the distance from the optical axis. The observed reflection spectra thus represent an average of reflection 
spectra taken for different optical densities. 

Finally, optical pumping between the ground state hyperfine levels and heating due to resonant absorption may occur while scanning the 
probe beam frequency and distort the spectra. These effects can however be accounted for in a simple rate equation model, which yields 
very good agreement with our observations.

\section{Discussion and conclusion}
\label{SecConclude}

The observations made with our apparatus clearly show features beyond Bragg scattering. These are due to two effects: Disorder arising 
from the thermal distribution of the atoms introduces strong absorption close to resonance, which limits the effective number of layers. 
%The disordered atoms also induce phase shifts, which give rise to asymmetries in the reflection profile. 
The second effect is multiple scattering between adjacent layers. Even though the above effects tend to distort or broaden the features, 
which are characteristic for multiple scattering, we find unambiguous signatures of multiple reflections. At this stage, it is however 
difficult to exactly quantify the number of layers involved in multiple scattering. In any case, for our present parameters the 
absorption spectra do not show any significant reduction at resonance, so that the qualification of \textit{photonic band gap} seems not 
adequate. 

An interesting question concerns the signature of atomic ordering in the transmitted and the absorbed light. In the thin grating regime, 
one would not expect that the atomic positions influence the absorption: the behavior of an absorbing atom does not depend on the 
location of the other atoms. Moreover, unlike for the reflection signal, interference plays no role in forward scattering nor in diffuse 
scattering. Hence there is no signature of atomic ordering in $T$ and $A$. The situation completely changes in the thick grating regime, 
where multiple scattering between subsequent atomic layers leads to interference between the light reflected from or transmitted through 
the layers. The globalization of the scattering process leads to interatomic correlation: Now it matters how the atoms are arranged. 
Multiple reflections give rise to stopping band gaps for certain ranges of light detuning or angle of incidence. It might therefore be 
more unambiguous to look for signatures of photonic band gaps in transmission or absorption spectra. 

\bigskip

To conclude we have shown that long-range spatial order in atomic clouds can have a dramatic influence on the scattering of light. 
We have extented earlier studies on Bragg scattering into the regime of thick gratings characterized by multiple reflections. Although 
signatures for reduced absorption could not be found due to the fatal influence of diffuse scattering, this represents a first step 
towards the realization of photonic band structures in optical lattices. 

Differently from photonic crystals or solid state systems, the scattering off optical lattices is weak except near atomic resonances. 
Therefore photonic stopping bands are expected to be very narrow. This bears the advantage that we can tune the optical density over a 
large range. However, the narrow resonance also implies that our system is intrinsically one-dimensional. An extrapolation to 
three-dimensional systems seems technically demanding, first of all because $3D$ optical lattices have low filling factors of typically 
$0.01$. There are however examples of lattices with unity filling factor \cite{DePue99}. Bose-condensates in the Mott insulator phase 
may prove useful to guarantee a high and regular occupation of the lattice sites \cite{Greiner02}. On the other hand, the sharpness of 
the resonance results in a very narrow tolerance angle for the stop band, which will make it difficult to obtain $3D$ PBGs. Coevorden 
\textit{et al.} \cite{Coevorden96} did numerical calculations of the band structure of a $3D$ optical lattice. To obtain a $3D$ PBG 
around an atomic transition of frequency $\omega$, they had to assume an excessively large spontaneous decay width, $\Gamma>0.01\omega$. 

A major advantage of using \textit{ultracold} atoms would be the total absence of diffuse scattering. Other possible technical upgrades 
include the use of standing waves with larger beam waists, thus extenting the radial size of the atomic layers, and the choice of smaller 
Bragg angles, which could be done by operating the dipole trap a few nm red-detuned from the same transition, the probe laser is tuned 
to. Thus, although it seems today difficult to compete with photonic crystals in terms of manipulating the propagation of light just by 
choosing a smart arrangement of gaseous atoms, there is much room left for improvements. 

\bigskip

We acknowledge financial support from the Landes\-stiftung Baden-W\"urt\-temberg.

\end{document}